\documentclass[12pt]{article} 

\usepackage{amsmath}  
\usepackage{amsfonts} 
\usepackage{graphicx} 
\usepackage{url}
\usepackage[bottom]{footmisc}
\usepackage{amssymb}
\usepackage{booktabs}
\usepackage[labelformat=simple]{subcaption}
\usepackage{textcomp}
\usepackage{mathrsfs}
\usepackage[shortlabels]{enumitem}
\usepackage{upgreek}
\usepackage{fullpage}
\usepackage[title]{appendix}
\usepackage[margin=1in]{geometry}
\usepackage{color}
\usepackage{times}
\usepackage{pgf,tikz}
\usepackage[raggedright]{titlesec}
\usepackage{multirow}
\usepackage{multicol}
\usepackage{array}
\usepackage{setspace}
\usepackage[format=plain,labelsep=period]{caption}
\usepackage{cite}
\usepackage{esint}
\usepackage{listings}

\titleformat{\section}{\bfseries}{\thesection}{1em}{}
\titleformat{\subsection}{\itshape}{\thesubsection}{1em}{}
\titleformat{\subsubsection}{}{\thesubsubsection}{1em}{}

\newcolumntype{L}[1]{>{\raggedright\let\newline\\\arraybackslash\hspace{0pt}}m{#1}}
\newcolumntype{C}[1]{>{\centering\let\newline\\\arraybackslash\hspace{0pt}}m{#1}}
\newcolumntype{R}[1]{>{\raggedleft\let\newline\\\arraybackslash\hspace{0pt}}m{#1}}

\begin{document}

\begin{table}[t]
\centering
\begin{tabular}{c}
{\large \textbf{Intrinsic variational structure of higher-derivative}} \\
{\large \textbf{formulations of classical mechanics}} \\ \\
{John W. Sanders} \\ \\
{Department of Mechanical Engineering} \\
{The Citadel, The Military College of South Carolina} \\
{171 Moutrie St, Charleston, SC 29409} \\
{jsande12@citadel.edu} \\
{ORCiD: 0000-0003-3059-3815} \\ \\
\end{tabular}
\end{table}

\section*{Abstract}

This paper investigates the geometric structure of higher-derivative formulations of classical mechanics. It is shown that every even-order formulation of classical mechanics higher than the second order is intrinsically variational, in the sense that the equations of motion are always derivable from a minimum action principle, even when the system is non-Hamiltonian. Particular emphasis is placed on the fourth-order formulation, as that is shown to be the lowest order for which the governing equations are intrinsically variational. The Noether symmetries and associated conservation laws of the fourth-order formulation, including its Hamiltonian, are derived along with the natural auxiliary conditions. The intrinsic variational structure of higher-derivative formulations makes it possible to treat non-Hamiltonian systems as if they were Hamiltonian, with immediate classical applications. A case study of the classical damped harmonic oscillator is presented for illustration, and an action is formulated for a higher-order Navier-Stokes equation. 



\section{Introduction}

Ever since the advent of classical mechanics, a clear line has been drawn between Hamiltonian and non-Hamiltonian systems~\cite{Hamilton1834,Hamilton1835}. However, a recent discovery~\cite{Sanders2021,Sanders2022,Sanders2023a,Sanders2023} has blurred that line. Evidently it is always possible to treat a non-Hamiltonian system as if it were Hamiltonian simply by considering the equations of motion at the fourth-order level in time. At the fourth-order level, the classical equations are intrinsically variational, in that they are always derivable from a stationary action integral, even when the system is non-conservative or non-holonomic~\cite{Sanders2023a}. This has already led to two classical applications: the direct modal analysis of damped dynamical systems~\cite{Sanders2022} and subsequently a new and more efficient algorithm for computing a damped system's resonant frequencies~\cite{Sanders2023}. In light of these recent applications, it is the present author's view that higher-derivative formulations of classical mechanics merit further investigation, and the central purpose of the present work is to investigate the geometric structure of such formulations. In fact, it will be shown that \emph{every} even-order formulation of classical mechanics higher than the second order is intrinsically variational.

The concept of a ``higher-derivative theory'' is not new and in fact dates back to the time of Ostrogradsky~\cite{Ostrogradsky1850}, who studied Lagrangians dependent on time derivatives higher than the first order. Ostrogradsky's celebrated instability theorem~\cite{Ostrogradsky1850} establishes that the Hamiltonian corresponding to such a Lagrangian is unbounded from below. Even so, higher-derivative Lagrangians have been studied extensively; see, for example, the work of VandenBerg and VanderVorst~\cite{VandenBerg2002}, Kalies and VanderVorst~\cite{Kalies2004}, and the references cited therein.

Ostrogradsky's work~\cite{Ostrogradsky1850} would later find relevance in quantum gravity physics~\cite{Pais1950,Bender2008,Smilga2009,Mostafazadeh2010,Baleanu2012,Chen2013}. It is well known that quantum mechanics is not easily unified with Einstein's general theory of relativity because the latter (a second-order theory, as the Einstein field equations are second-order in the metric tensor via the Ricci curvature) is non-renormalizable~\cite{Pais1950}. Higher-derivative theories have been considered as candidate theories of quantum gravity because they are renormalizable~\cite{Mostafazadeh2010}. Unfortunately, consistent with Ostrogradsky's instability theorem~\cite{Ostrogradsky1850}, such theories usually admit so-called ``bad ghosts'': quantum states with negative kinetic energies, resulting in the loss of unitarity~\cite{Bender2008}. One notable exception is the fourth-order Pais-Uhlenbeck oscillator~\cite{Pais1950,Bender2008}. Still, quantum gravity remains an open problem, and it is not clear whether higher-derivative theories hold the key to its resolution.

It was not until recently that higher-derivative equations found applications to classical mechanics and engineering. To be sure, higher-order \emph{derivatives} have been of classical interest for quite some time. One very well known example is the evolution of the vertical roller coaster loop from perfectly circular to a clothoid shape in order to avoid discontinuities in jerk~\cite{Tipler2007}. Another notable example is the design of control schemes to achieve minimum-snap trajectories for unmanned aerial vehicles; see the work of Mellinger and Kumar~\cite{Mellinger2011} and citing papers. However, the classical \emph{equations of motion} have always been second-order in time, and it would seem that equations involving higher-order time-derivatives have not been given serious consideration by classical mechanicians since Ostrogradsky~\cite{Ostrogradsky1850}. 

Sanders~\cite{Sanders2021} (the present author) showed that the damped harmonic oscillator belongs to the family of fourth-order Pais-Uhlenbeck oscillators~\cite{Pais1950,Bender2008}, with the corollary that a single damped oscillator is mathematically equivalent to two coupled ``dual oscillators'' with no dampers but with springs of complex-valued stiffnesses~\cite{Sanders2021}. By extending that result~\cite{Sanders2021}, Sanders~\cite{Sanders2022} was able to perform direct modal analysis of non-proportionally damped linear systems with arbitrary degrees of freedom, and at the same time opened the door to direct modal analysis of damped nonlinear oscillators with viscous damping and power-law hardening~\cite{Sanders2022}. Sanders and Inman~\cite{Sanders2023} then showed that the resonant frequency computation provided by the dual oscillator approach is generally more efficient than the traditional method of Foss~\cite{Foss1956} and Traill-Nash~\cite{TraillNash1983}.

Sanders~\cite{Sanders2022} recognized that the fourth-order dynamics were intrinsically variational and subsequently established a general variational formulation for the fourth-order equations of motion for any mechanical system~\cite{Sanders2023a}. Sanders~\cite{Sanders2023a} considered an action of the form
\begin{equation}\label{eq:action}
\mathcal{S}^{*}[q_{i}(t)]=\int_{t_{1}}^{t_{2}}\frac{1}{2}\mathcal{R}_{i}\mathcal{R}_{i}dt, \quad \mathcal{R}_{i}\equiv\ddot{q}_{i}-f_{i}(q_{j},\dot{q}_{j},t), \quad i,j=1,2,...,n,
\end{equation}
where the $q_{i}=q_{i}(t)$ are suitable generalized coordinates\footnote{In order to maintain dimensional consistency, the generalized coordinates may require nondimensionalization.}, $t_{1}$ and $t_{2}$ are arbitrary times, the $\mathcal{R}_{i}$ are the residuals of the second-order formulation, $n$ is the system's degree of freedom, and we employ the Einstein summation convention on repeated subscript indices. As noted by Sanders~\cite{Sanders2023a}, the ``Lagrangian'' here,
\begin{equation}\label{eq:Lagrangian}
L^{*}(t,q_{i},\dot{q}_{i},\ddot{q}_{i})
=\frac{1}{2}\mathcal{R}_{i}\mathcal{R}_{i},
\end{equation}
bears a superficial resemblance to Gauss's constraint function~\cite{Gauss1829,Lanczos1970}, although the two are not generally identical. Varying the coordinates $q_{i}(t)$, the first variation $\delta L^{*}=\mathcal{R}_{i}\delta\mathcal{R}_{i}
$ vanishes for the actual motion satisfying $\mathcal{R}_{i}=0$, and the second variation $\delta^{2} L^{*}=\delta\mathcal{R}_{i}\delta\mathcal{R}_{i}$ is strictly non-negative. Thus, it is clear that $L^{*}$ achieves a local minimum for the actual motion, as does $\mathcal{S}^{*}$. The minimization of $L^{*}$ is the well-known principle of least squares in numerical analysis. The minimization of $\mathcal{S}^{*}$---which is simply $L^{*}$ averaged over an arbitrary time interval---might be called the ``time-averaged principle of least squares.''

We pause here to note that the Lagrangian $L^{*}$ in \eqref{eq:Lagrangian} is not the \emph{only} Lagrangian that gives rise to the fourth-order equations. Sometimes the quantity $\mathcal{R}_{i}\mathcal{R}_{i}$ contains terms that can be expressed as a total time derivative, in which case it is well known that those terms can be eliminated from the action without affecting the equations of motion. We will return to this issue in Section~\ref{sec:DHO}.

The significance of this $\mathcal{S}^{*}$ is that it provides a stationary action for the fourth-order equations of any classical mechanical system. Unlike the ordinary Hamilton action~\cite{Hamilton1834,Hamilton1835}, $\mathcal{S}^{*}$ applies even to non-conservative and non-holonomic systems, blurring the distinction between Hamiltonian and non-Hamiltonian systems and allowing non-Hamiltonian systems to be treated as if they were Hamiltonian. The existence of practical applications for the fourth-order formulation warrants further investigation of higher-order formulations of classical mechanics. 

The remainder of this paper is organized as follows. Section~\ref{sec:structure} provides a general exposition of the intrinsic variational structure of the fourth-order formulation, including the various Noether symmetries, associated conservation laws, and natural auxiliary conditions. In Section~\ref{sec:DHO}, the classical damped harmonic oscillator is used as a case study to illustrate the general results of Section~\ref{sec:structure}. Section~\ref{sec:arbhighorders} generalizes the results of Section~\ref{sec:structure} to arbitrarily high even-order formulations. Finally, Section~\ref{sec:conclusion} concludes with a brief summary of the main results and a candidate action for a higher-order Navier-Stokes equation~\cite{Stokes1845}.

\section{Variational structure of the fourth-order formulation}\label{sec:structure}

The first variation of $\mathcal{S}^{*}$ is
\begin{equation}
\delta\mathcal{S}^{*}=\int_{t_{1}}^{t_{2}}\left(\mathcal{R}_{i}\delta\ddot{q}_{i}-\mathcal{R}_{j}\frac{\partial f_{j}}{\partial \dot{q}_{i}}\delta{\dot{q}}_{i}-\mathcal{R}_{j}\frac{\partial f_{j}}{\partial q_{i}}\delta{q}_{i}\right)dt,
\end{equation}
and upon integration by parts we have
\begin{equation}\label{eq:deltaSstar}
\delta\mathcal{S}^{*}=\int_{t_{1}}^{t_{2}}\left[\ddot{\mathcal{R}}_{i}+\frac{d}{dt}\left(\mathcal{R}_{j}\frac{\partial f_{j}}{\partial \dot{q}_{i}}\right)-\mathcal{R}_{j}\frac{\partial f_{j}}{\partial q_{i}}\right]\delta{q}_{i}dt+\left[-\left(\mathcal{R}_{j}\frac{\partial f_{j}}{\partial \dot{q}_{i}}+\dot{\mathcal{R}}_{i}\right)\delta{q}_{i}+\mathcal{R}_{i}\delta\dot{q}_{i}\right]_{t_{1}}^{t_{2}}.
\end{equation}
When viewed as a principle of motion, the minimization of $\mathcal{S}^{*}$ yields the Euler-Lagrange equations
\begin{equation}\label{eq:EL}
\ddot{\mathcal{R}}_{i}+\dot{\mathcal{R}}_{j}\frac{\partial f_{j}}{\partial \dot{q}_{i}}+\mathcal{R}_{j}\left(\frac{d}{dt}\frac{\partial f_{j}}{\partial \dot{q}_{i}}-\frac{\partial f_{j}}{\partial q_{i}}\right)=0.
\end{equation} 
These equations are fourth-order in time, and are mathematically equivalent to the original second-order equations $\mathcal{R}_{i}=0$ with identical initial conditions on $q_{i}$, $\dot{q}_{i}$, $\ddot{q}_{i}$, and $\dddot{q}_{i}$. Specifically, if $q_{i}(0)=q^{0}_{i}$ and $\dot{q}_{i}(0)=v^{0}_{i}$, then we require
\begin{equation}\label{eq:IC1}
\ddot{q}_{i}(0)=f_{i}(q^{0}_{j},v^{0}_{j},0)
\end{equation}
and
\begin{equation}\label{eq:IC2}
\dddot{q}_{i}(0)=\dot{f}_{i}(q^{0}_{j},v^{0}_{j},0)
\end{equation}
to recover the solution to the original second-order problem.

\subsection{Noether symmetries and conservation laws}\label{sec:Noether}

As is well known, every continuous symmetry in a Lagrangian has an associated conservation law~\cite{Noether1918}, and the present formulation is no exception. Here we will investigate the symmetries and associated conserved quantities for the Lagrangian $L^{*}$ given by \eqref{eq:Lagrangian}.

Corresponding to each of the coordinates $q_{i}$ and the associated velocities $v_{i}=\dot{q}_{i}$ are canonically conjugate momenta, which can be identified directly from the boundary term in~\eqref{eq:deltaSstar}. Specifically, the momenta conjugate the coordinates $q_{i}$ are
\begin{equation}\label{eq:pi}
p_{i}=-\left(\mathcal{R}_{j}\frac{\partial f_{j}}{\partial \dot{q}_{i}}+\dot{\mathcal{R}}_{i}\right),
\end{equation}
and the momenta conjugate to the velocities $v_{i}$ are
\begin{equation}\label{eq:Pi}
P_{i}=\mathcal{R}_{i}.
\end{equation}
Clearly these momenta vanish identically for the actual motion (${\mathcal{R}}_{i}=0$) and are thus trivially constants of the motion. Nevertheless, they carry geometric significance within the context of the complete set of solutions to the fourth order equations. When $L^{*}$ exhibits symmetry in one of the generalized coordinates $q_{i}$, the conjugate momentum $p_{i}$ is conserved for \emph{all} solutions to the fourth-order equations of motion---not only the true solution but the extraneous solutions as well. Furthermore, since $L^{*}$ is not unique, neither are the generalized momenta. Eliminating total time derivatives from $L^{*}$ will lead to different generalized momenta, which may not vanish identically for the actual motion. We will return to that issue in Section~\ref{sec:DHO}.

Just as symmetry in a coordinate leads to conservation of the conjugate momentum, temporal symmetry is associated with the conservation of a suitable Hamiltonian. The Hamiltonian associated with $L^{*}$ is obtained via the Legendre transform
\begin{equation}\label{eq:Hstar}
H^{*}=p_{i}v_{i}+P_{i}\dot{v}_{i}-L^{*}=\frac{1}{2}P_{i}P_{i}+P_{i}f_{i}(q_{j},v_{j},t)+p_{i}v_{i},
\end{equation}
where $L^{*}=\tfrac{1}{2}\mathcal{R}_{i}\mathcal{R}_{i}=\tfrac{1}{2}P_{i}P_{i}$ and 
\begin{equation}\label{eq:vidot}
\dot{v}_{i}=\ddot{q}_{i}=\mathcal{R}_{i}+f_{i}(q_{j},v_{j},t)=P_{i}+f_{i}(q_{j},v_{j},t)
\end{equation}
have been expressed in terms of the coordinates, velocities, and conjugate momenta. Clearly this Hamiltonian is very different from the total mechanical energy of the system, although it is still a conserved quantity. Indeed, like the conjugate momenta, $H^{*}$ vanishes for the actual motion ($\mathcal{R}_{i}=0$) and is thus a trivial constant of the motion. Nevertheless, it carries geometric significance within the context of the complete set of solutions to the fourth order equations. When $L^{*}$ exhibits time symmetry, $H^{*}$ is conserved for \emph{all} solutions to the fourth-order equations of motion---not just the true solution but also the extraneous solutions. 

\subsection{Hamilton's equations}

To obtain Hamilton's canonical equations, we express $\mathcal{S}^{*}$ in terms of $H^{*}$,
\begin{equation}\label{eq:action2}
\mathcal{S}^{*}=\int_{t_{1}}^{t_{2}}\left(p_{i}\dot{q}_{i}+P_{i}\dot{v}_{i}-H^{*}\right)\text{d}t,
\end{equation}
evaluate the first variation, and integrate by parts:
\begin{align}
\delta\mathcal{S}^{*}=\int_{t_{1}}^{t_{2}}\biggl[\biggr.&\left(\dot{q}_{i}-\frac{\partial H^{*}}{\partial p_{i}}\right)\delta p_{i}-\left(\dot{p}_{i}+\frac{\partial H^{*}}{\partial q_{i}}\right)\delta q_{i} \\
&\biggl.+\left(\dot{v}_{i}-\frac{\partial H^{*}}{\partial P_{i}}\right)\delta P_{i}-\left(\dot{P}_{i}+\frac{\partial H^{*}}{\partial v_{i}}\right)\delta v_{i}\biggr]\text{d}t.
\end{align}
Setting $\delta\mathcal{S}^{*}=0$ yields the following canonical equations:
\begin{align}
&\dot{q}_{i}=\frac{\partial H^{*}}{\partial p_{i}}, &\dot{v}_{i}&=\frac{\partial H^{*}}{\partial P_{i}}\\
&\dot{p}_{i}=-\frac{\partial H^{*}}{\partial q_{i}}, &\dot{P}_{i}&=-\frac{\partial H^{*}}{\partial v_{i}}.
\end{align}
The first set of equations, $\dot{q}_{i}={\partial H^{*}}/{\partial p_{i}}$, recovers $\dot{q}_{i}=v_{i}$. The second, $\dot{v}_{i}={\partial H^{*}}/{\partial P_{i}}$, recovers the righthand side of~\eqref{eq:vidot}, which is mathematically equivalent to~\eqref{eq:Pi}. The third, $\dot{p}_{i}=-{\partial H^{*}}/{\partial q_{i}}$, recovers the Euler-Lagrange equations~\eqref{eq:EL}. The fourth and last set of equations, $\dot{P}_{i}=-{\partial H^{*}}/{\partial v_{i}}$, recovers~\eqref{eq:pi}. These canonical equations are therefore mathematically equivalent to the fourth-order problem, which is in turn mathematically equivalent to the original second-order problem. 

\subsection{Hamilton-Jacobi theory}

The Hamiltonian formalism reaches its pinnacle in the transformation theory of Hamilton and Jacobi, which is encapsulated in the Hamilton-Jacobi equation~\cite{Hamilton1833,Hamilton1834,Hamilton1835,Jacobi1837,Jacobi18421843,Whittaker1904,Lanczos1970}. The merit of the Hamilton-Jacobi equation is that reduces the problem of finding the (generally multitudinous) generalized coordinates $q_{i}(t)$ and velocities $v_{i}(t)$ to that of finding a \emph{single scalar function}: Hamilton's principal function $\text{S}^{*}=\text{S}^{*}(q_{i},v_{j},t)$. This function serves as the generating function for a canonical transformation to a new set of coordinates and conjugate momenta for which the Hamiltonian $H^{*}$ vanishes \emph{identically} (\emph{i.e.},~not just for the actual motion, but for every conceivable motion), rendering Hamilton's equations trivial: the new coordinates and their conjugate momenta are simply equal to their initial values. Thus, $\text{S}^{*}$ contains within it the entire solution to the initial value problem, and to have found the former is to have solved the latter.

Historically, the Hamilton-Jacobi theory is celebrated for unifying particle mechanics with wave optics~\cite{Hamilton1833} and for providing the foundation of quantum mechanics in the Schr\"{o}dinger equation~\cite{Schroedinger1926,Schroedinger1926a}. Until now, the Hamilton-Jacobi theory has only applied to Hamiltonian systems. We will now obtain a Hamilton-Jacobi equation valid even for non-Hamiltonian systems.

For the purposes of deriving the Hamilton-Jacobi equation, Hamilton's principal function $\text{S}^{*}=\text{S}^{*}(q_{i},v_{j},t)$ may be defined as the action integral $\mathcal{S}^{*}$, evaluated for generalized coordinates and velocities satisfying the Euler-Lagrange equations~\eqref{eq:EL}, and with the time integral carried out from an arbitrary initial time $t_{0}$ to the current (variable) time $t$:
\begin{equation}\label{eq:HPF}
\text{S}^{*}(q_{i},v_{j},t)=\int_{t_{0}}^{t}\frac{1}{2}\tilde{\mathcal{R}}_{i}\tilde{\mathcal{R}}_{i}\text{d}t,
\end{equation}
where $\tilde{\mathcal{R}}_{i}$ denotes the residual ${\mathcal{R}}_{i}$ evaluated for generalized coordinates and velocities satisfying the Euler-Lagrange equations~\eqref{eq:EL}. 

Starting from~\eqref{eq:HPF} and evaluating the first variation of $\text{S}^{*}$ as we did with the action $\mathcal{S}^{*}$, we find that
\begin{equation}
\delta\text{S}^{*}=\left[p_{i}\delta{q}_{i}+P_{i}\delta v_{i}\right]^{t}_{t_{0}},
\end{equation}
where we have used the fact that the Euler-Lagrange equations~\eqref{eq:EL} are satisfied. But, by the chain rule,
\begin{equation}
\delta\text{S}^{*}=\frac{\partial\text{S}^{*}}{\partial q_{i}}\delta q_{i}+\frac{\partial\text{S}^{*}}{\partial v_{i}}\delta v_{i}.
\end{equation}
We may thus identify the conjugate momenta with the partial derivatives of $\text{S}^{*}$:
\begin{equation}\label{eq:momenta}
p_{i}=\frac{\partial\text{S}^{*}}{\partial q_{i}}, \quad P_{i}=\frac{\partial\text{S}^{*}}{\partial v_{i}}.
\end{equation}

Now starting from~\eqref{eq:HPF} and evaluating the total time derivative, we have
\begin{equation}
\frac{\text{d}\text{S}^{*}}{\text{d}t}=\frac{1}{2}\tilde{\mathcal{R}}_{i}\tilde{\mathcal{R}}_{i}.
\end{equation}
But, again by the chain rule,
\begin{equation}
\frac{\text{d}\text{S}^{*}}{\text{d}t}=\frac{\partial\text{S}^{*}}{\partial q_{i}}\dot{q}_{i}+\frac{\partial\text{S}^{*}}{\partial v_{i}}\dot{v}_{i}+\frac{\partial\text{S}^{*}}{\partial t}.
\end{equation}
We therefore have that
\begin{equation}
\frac{\partial\text{S}^{*}}{\partial q_{i}}\dot{q}_{i}+\frac{\partial\text{S}^{*}}{\partial v_{i}}\dot{v}_{i}-\frac{1}{2}\tilde{\mathcal{R}}_{i}\tilde{\mathcal{R}}_{i}+\frac{\partial\text{S}^{*}}{\partial t}=0.
\end{equation}
All of the terms on the lefthand side \emph{except} for the partial time derivative represent the Hamiltonian $H^{*}$, with the conjugate momenta replaced by the corresponding partial derivatives according to~\eqref{eq:momenta}. Replacing the conjugate momenta accordingly in~\eqref{eq:Hstar}, we arrive at the Hamilton-Jacobi equation
\begin{equation}
\frac{1}{2}\frac{\partial\text{S}^{*}}{\partial v_{i}}\frac{\partial\text{S}^{*}}{\partial v_{i}}+\frac{\partial\text{S}^{*}}{\partial v_{i}}f_{i}(q_{j},v_{j},t)+\frac{\partial\text{S}^{*}}{\partial q_{i}}v_{i}+\frac{\partial\text{S}^{*}}{\partial t}=0.
\end{equation}
Unlike the classical Hamilton-Jacobi equation, the above is valid for all mechanical systems, both Hamiltonian and non-Hamiltonian. It reduces every single problem in classical mechanics to the search for a single scalar function, $\text{S}^{*}(q_{i},v_{j},t)$.

\subsection{Natural auxiliary conditions}\label{sec:natACs}

Another beautiful feature of variational principles is their ability to furnish natural auxiliary conditions in addition to the equations of motion. Examining the boundary terms in~\eqref{eq:deltaSstar},
\begin{equation}
\left[p_{i}\delta{q}_{i}+P_{i}\delta\dot{q}_{i}\right]_{t_{1}}^{t_{2}},
\end{equation}
the natural auxiliary conditions would be that $p_{i}=0$ and $P_{i}=0$ at both $t=t_{1}$ and $t=t_{2}$ (or equivalently, that $\mathcal{R}_{i}=0$ and $\dot{\mathcal{R}}_{i}=0$ at both $t=t_{1}$ and $t=t_{2}$). Recall that initial conditions making ${\mathcal{R}}_{i}=0$ and $\dot{\mathcal{R}}_{i}=0$ at time $t=0$ serve to guarantee that the solution to the fourth-order formulation agrees with the solution to the original, second-order formulation. Thus it may be said that the actual motion is the ``natural evolution'' of the fourth-order system resulting from the action $\mathcal{S}^{*}$ as given by~\eqref{eq:action}.

\subsection{Generalized stiffnesses and damping coefficients}\label{sec:generalizedLandH}

We pause here to note the superficial resemblance of the governing fourth-order equations \eqref{eq:EL} to a system of damped oscillators. This suggests defining generalized stiffnesses
\begin{equation}\label{eq:genL}
\ell_{ji}\equiv\frac{d}{dt}\frac{\partial f_{j}}{\partial \dot{q}_{i}}-\frac{\partial f_{j}}{\partial q_{i}}
\end{equation}
and generalized damping coefficients
\begin{equation}\label{eq:genH}
h_{ji}\equiv-\frac{\partial f_{j}}{\partial \dot{q}_{i}},
\end{equation}
so that \eqref{eq:EL} reads
\begin{equation}
\ddot{\mathcal{R}}_{i}-\dot{\mathcal{R}}_{j}h_{ji}+\mathcal{R}_{j}\ell_{ji}=0.
\end{equation}
For reference, the second-order system $\ddot{q}_{i}+H_{ij}\dot{q}_{j}+L_{ij}q_{j}=0$ would have generalized damping coefficients $h_{ij}=H_{ij}$ and generalized stiffnesses $\ell_{ij}=L_{ij}$. In effect, the residuals ${\mathcal{R}}_{i}$---which vanish identically for the actual motion---more generally exhibit a sort of anti-damped oscillation in the broader context of the complete set of solutions to the fourth-order equations.

The meaning of the generalized damping coefficients is clear enough: $h_{ji}=0$ precisely when $f_{j}$ does not depend on $\dot{q}_{i}$, in which case the governing equation for $\mathcal{R}_{i}$ does not depend explicitly on $\dot{\mathcal{R}}_{j}$. If all of the $h_{ji}$ vanish, then the system is completely undamped. 

The meaning of the generalized stiffnesses is less clear but more interesting. Evidently $\ell_{ji}=0$ precisely when $f_{j}$ is stationary with respect to small variations in $q_{i}(t)$, in which case the governing equation for $\mathcal{R}_{i}$ does not depend explicitly on ${\mathcal{R}}_{j}$. The present author is not aware of a classical interpretation for this scenario, but it may be that it has some practical implication for the actual motion.

\section{Case study: the classical damped harmonic oscillator}\label{sec:DHO}

The classical damped harmonic osillator has residual
\begin{equation}
\mathcal{R}\equiv\ddot{x}+2\omega\zeta\dot{x}+\omega^{2}x,
\end{equation}
where $\omega>0$ is the undamped natural frequency and $\zeta\geq0$ is the damping ratio. Nondimensionalization can be achieved by setting $\omega=1$, although because there is only one coordinate here nondimensionalization is not necessary. For reference, the generalized damping coefficient~\eqref{eq:genH} is $h=2\omega\zeta$ and the generalized stiffness~\eqref{eq:genL} is $\ell=\omega^{2}$.

The Lagrangian $L^{*}$, as given by \eqref{eq:Lagrangian}, is
\begin{equation}\label{eq:LagrangianDHO}
L^{*}=\frac{1}{2}\left(\ddot{x}^{2}+4\omega\zeta\ddot{x}\dot{x}+2\omega^{2}\ddot{x}x+4\omega^{2}\zeta^{2}\dot{x}^{2}+4\omega^{3}\zeta\dot{x}x+\omega^{4}x^{2}\right),
\end{equation}
with resulting equation of motion
\begin{equation}\label{eq:4ODHO}
\ddddot{x}+4\omega^{2}\left(\frac{1}{2}-\zeta^{2}\right)\ddot{x}+\omega^{4}x=0,
\end{equation}
as was previously obtained by Sanders~\cite{Sanders2021,Sanders2022}, who noted that the same equation could be obtained by first differentiating the second-order equation twice and manually eliminating odd-order derivatives (damping terms). The conjugate momentum associated with $x$ is
\begin{equation}
p=4\omega^{2}\zeta^{2}\dot{x}+2\omega^{3}\zeta{x}-\dddot{x}-\omega^{2}\dot{x},
\end{equation}
and the Hamiltonian is
\begin{equation}\label{eq:HamiltonianDHO}
H^{*}=\frac{1}{2}\left[\ddot{x}^{2}-4\omega^{2}\left(\frac{1}{2}-\zeta^{2}\right)\dot{x}^{2}-\omega^{4}x^{2}\right]-\dddot{x}\dot{x},
\end{equation}
both of which can easily be shown to vanish for the actual motion.

Recall that the Lagrangian is not unique. Two of the terms in~\eqref{eq:LagrangianDHO} can be eliminated as they are expressible as total time derivatives: $\ddot{x}\dot{x}\sim d/dt(\dot{x}^{2})$, $\dot{x}x\sim d/dt(x^{2})$. This results in a different Lagrangian,
\begin{equation}\label{eq:LagrangianDHO2}
{L}^{*}_{(2)}=\frac{1}{2}\left(\ddot{x}^{2}+2\omega^{2}\ddot{x}x+4\omega^{2}\zeta^{2}\dot{x}^{2}+\omega^{4}x^{2}\right),
\end{equation}
which yields the same equation of motion~\eqref{eq:4ODHO} and the same Hamiltonian~\eqref{eq:HamiltonianDHO} but a different conjugate momentum
\begin{equation}
{p}_{(2)}=4\omega^{2}\zeta^{2}\dot{x}-\dddot{x}-\omega^{2}\dot{x},
\end{equation}
which differs from $p$ by $2\omega^{3}\zeta{x}$ and therefore does \emph{not} vanish identically for the actual motion. Additionally, the term $\ddot{x}{x}$ in~\eqref{eq:LagrangianDHO} may be integrated by parts once, yielding yet another Lagrangian
\begin{equation}\label{eq:LagrangianDHO3}
{L}^{*}_{(3)}=\frac{1}{2}\left[\ddot{x}^{2}-4\omega^{2}\left(\frac{1}{2}-\zeta^{2}\right)\dot{x}^{2}+\omega^{4}x^{2}\right],
\end{equation}
which is the form of the Lagrangian considered by Sanders~\cite{Sanders2021,Sanders2022}. It yields the same equation of motion~\eqref{eq:4ODHO} and the same Hamiltonian~\eqref{eq:HamiltonianDHO} but yet another conjugate momentum
\begin{equation}
{p}_{(3)}=-4\omega^{2}\left(\frac{1}{2}-\zeta^{2}\right)\dot{x}-\dddot{x},
\end{equation}
which differs from $p_{(2)}$ by $\omega^{2}\dot{x}$ and does {not} vanish identically for the actual motion either. The non-uniqueness of the Lagrangian (and thus also the generalized momenta) is a well known phenomenon and does not represent a deficiency in the formulation. The family of all such Lagrangians will yield the same equation of motion~\eqref{eq:4ODHO}.

The fact that the governing equation~\eqref{eq:4ODHO} has no odd-order derivatives was exploited by Sanders and Inman~\cite{Sanders2023} to perform rapid computations of resonant frequencies for non-proportionally damped systems. By writing the fourth-order system as an augmented second-order system, Sanders and Inman~\cite{Sanders2023} obtained the resonant frequencies with fewer computations than the previous standard approach~\cite{Foss1956,TraillNash1983}. This was a direct application of the intrinsic variational structure of the fourth-order formulation. Indeed, more generally, this intrinsic variational structure allows us to treat any non-Hamiltonian system as if it were Hamiltonian.

\section{Generalization to higher orders}\label{sec:arbhighorders}

The same strategy can be generalized to construct variational formulations for arbitrarily high even-order formulations of classical mechanics. Taking $N$ time derivatives of the residuals $\mathcal{R}_{i}$ of the second-order formulation, we obtain
\begin{equation}
\mathcal{R}_{i}^{(N)}={q}_{i}^{(2+N)}-f_{i}^{(N)}(q_{j},\dot{q}_{j},t)=0, \quad i,j=1,2,...,n,
\end{equation}
where a parenthetical superscipt denotes a number of time derivatives: $(\cdot)^{(N)}\equiv d^{N}/dt^{N}(\cdot)$ and $N\geq0$. Now consider the following action:
\begin{equation}\label{eq:arbhighaction}
\mathcal{S}^{*}_{N}[q_{i}(t)]=\int_{t_{1}}^{t_{2}}\frac{1}{2}\mathcal{R}_{i}^{(N)}\mathcal{R}_{i}^{(N)}dt.
\end{equation}
Varying the generalized coordinates $q_{i}(t)$, the first variation is 
\begin{equation}
\delta\mathcal{S}^{*}_{N}=\int_{t_{1}}^{t_{2}}\mathcal{R}_{i}^{(N)}\delta\mathcal{R}_{i}^{(N)}dt=\int_{t_{1}}^{t_{2}}\left[\mathcal{R}_{i}^{(N)}\delta{q}_{i}^{(2+N)}-\mathcal{R}_{i}^{(N)}\delta f_{i}^{(N)}\right]dt,
\end{equation}
which clearly vanishes for the actual motion satisfying $\mathcal{R}_{i}\equiv0$. Upon integration by parts,
\begin{equation}
\delta\mathcal{S}^{*}_{N}=\int_{t_{1}}^{t_{2}}\left[(-1)^{2+N}\mathcal{R}_{i}^{(2N+2)}\delta{q}_{i}-\mathcal{R}_{i}^{(N)}\delta f_{i}^{(N)}\right]dt.
\end{equation}
For sufficiently smooth $f_{i}$, the equations of motion obtained by setting $\delta\mathcal{S}^{*}_{N}=0$ for arbitrary $\delta{q}_{i}$ will be mathematically equivalent to $\mathcal{R}_{i}^{(2N+2)}=0$ for motions satisfying $\mathcal{R}_{i}\equiv0$. But $\mathcal{R}_{i}^{(2N+2)}=0$ are just the $(2N+4)$th-order equations of motion. With $N=0$ we obtain the fourth-order equations, with $N=1$ we obtain the sixth-order equations, and so on.

Furthermore, it is clear that $\mathcal{S}^{*}_{N}$ achieves a local minimum for the actual motion, since
\begin{equation}
\delta^{2}\mathcal{S}^{*}_{N}=\int_{t_{1}}^{t_{2}}\delta\mathcal{R}_{i}^{(N)}\delta\mathcal{R}_{i}^{(N)}dt
\end{equation}
is strictly non-negative.

We conclude that every even-order formulation of classical mechanics higher than the second order is intrinsically variational, in the sense that the equations of motion are derivable from a minimum action principle even when the system is non-Hamiltonian. And since it is well known that no odd-order equation has variational structure~\cite{Lanczos1970}, it follows that the fourth order is the lowest order for which the classical equations are intrinsically variational.

\section{Summary and conclusion}\label{sec:conclusion}

We may summarize the original results of the present work as follows:
\begin{enumerate}[1.]
  \item Every even-order formulation of classical mechanics higher than the second order is intrinsically variational, even for non-Hamiltonian systems. A suitable action for the $(2N+4)$th-order formulation is provided by~\eqref{eq:arbhighaction}, with $\mathcal{R}_{i}$ the residuals of the second-order formulation.
  \item The fourth-order formulation is the lowest-order formulation that is intrinsically variational. Various symmetries and associated conservation laws are derived in Section~\ref{sec:structure}. Although the Lagrangian, and therefore the generalized momenta, are not unique, the family of all such Lagrangians yields the same governing equations~\eqref{eq:EL}. The Hamiltonian~\eqref{eq:Hstar} vanishes identically for the actual motion.
  \item The natural auxiliary conditions of the fourth-order formulation, when enforced, give rise to motion consistent with the second-order formulation. Thus it may be said that the actual motion is the natural evolution of the fourth-order formulation.
\end{enumerate}
A direct consequence of these results is that it is always possible to treat a non-Hamiltonian system as Hamiltonian by considering higher-order equations of motion. For instance, a suitable action for a higher-order Navier-Stokes equation~\cite{Stokes1845} is given by
\begin{equation}\label{eq:NSaction}
\mathcal{S}^{*}=\int\frac{1}{2}\mathcal{R}_{i}\mathcal{R}_{i}dx^{4}, \quad \mathcal{R}_{i}\equiv\rho\frac{dv_{i}}{dt}+\frac{\partial p}{\partial x_{i}}-\mu\frac{\partial^{2}v_{i}}{\partial x_{j}\partial x_{j}}-(\mu+\lambda)\frac{\partial^{2}v_{j}}{\partial x_{j}\partial x_{i}}-b_{i},
\end{equation}
where the integral is carried out over the spacetime occupied by the fluid and the $\mathcal{R}_{i}$ are the residuals of the second-order Navier-Stokes equation~\cite{Stokes1845}, given here in terms of the density field $\rho(x_{j},t)$, velocity field $v_{i}(x_{j},t)$, pressure field $p(x_{j},t)$, body force field $b_{i}(x_{j},t)$, and viscosity coefficients $\mu$, $\lambda$. What other applications such a formulation may have are yet to be seen. It may be that questions of existence and uniqueness of solutions can be addressed by appealing to the properties of a suitable action~\cite{Kalies2004}.

%
%
%
%
%
%

%
%
%
%

\begin{spacing}{0.25}
\small
\bibliographystyle{asmems4}
\bibliography{Sanders2023arXivR2}
\end{spacing}

\end{document}